\documentclass[preprint,11pt]{aastex}
\usepackage{pslatex,times,natbib,emulateapj5}

\renewcommand{\arraystretch}{.6}
\newcommand{\civ}{C\,{\sc iv}}

\newcommand{\hb}{H$\beta$}

\newcommand{\kms}{km\,s$^{-1}$}

\newcommand{\mgii}{Mg\,{\sc ii}}

\newcommand{\oiii}{[O\,{\sc iii}]}

\newcommand{\cii}{C\,{\sc ii}}

\newcommand{\ciii}{C\,{\sc iii}]}
\newcommand{\Ciii}{C\,{\sc iii}}

\newcommand{\CIV}{C\,{\sc iv}\,$\lambda\lambda$1548,1550}

\newcommand{\crii}{Cr\,{\sc ii}}

\newcommand{\feii}{Fe\,{\sc ii}}

\newcommand{\lya}{Ly$\alpha$}

\newcommand{\lyb}{Ly$\beta$}

\newcommand{\mgi}{Mg\,{\sc i}}

\newcommand{\Nv}{N\,{\sc v}}
\newcommand{\NV}{N\,{\sc v}\,$\lambda\lambda$1238,1242}

\newcommand{\oi}{O\,{\sc i}}

\newcommand{\ovi}{O\,{\sc vi}}

\newcommand{\SIii}{Si\,{\sc ii}}
\newcommand{\SIiii}{Si\,{\sc iii}}

\newcommand{\SIiv}{Si\,{\sc iv}}

\newcommand{\znii}{Zn\,{\sc ii}}

\newcommand{\fl}{$F_{\lambda}$}
\newcommand{\nlya}{SDSS J0952+0114}
\newcommand{\lyg}{Ly$\gamma$}

\shortauthors{Hall et al.}
\shorttitle{Missing Broad \lya}
\journalinfo{SDSS Publication \#269 --- Submitted to AJ Feb. 9, 2004; Accepted May 12, 2004}
\submitted{}
\slugcomment{}

\begin{document}

\title{A Quasar Without Broad Ly$\alpha$ Emission} 

\author{
Patrick B. Hall,\altaffilmark{1,2}
Stephanie A. Snedden,\altaffilmark{3}
Martin Niederste-Ostholt,\altaffilmark{1}
Daniel J. Eisenstein,\altaffilmark{4}
Michael A. Strauss,\altaffilmark{1}
Donald G. York,\altaffilmark{5}
Donald P. Schneider\altaffilmark{6}
}
\altaffiltext{1}{Princeton University Observatory, Princeton, NJ 08544-1001}
\altaffiltext{2}{Departamento de Astronom\'{\i}a y Astrof\'{\i}sica, Facultad de
F\'{\i}sica, Pontificia Universidad Cat\'{o}lica de Chile, Casilla 306, Santiago
22, Chile; E-mail: phall@astro.puc.cl}
\altaffiltext{3}{Apache Point Observatory, P.O. Box 59, Sunspot, NM 88349-0059}
\altaffiltext{4}{Steward Observatory, The University of Arizona,
933 North Cherry Avenue, Tucson AZ 85721}
\altaffiltext{5}{Department of Astronomy and Astrophysics and Enrico Fermi
Institute, The University of Chicago, 5640 S. Ellis Ave., Chicago, IL 60637}
\altaffiltext{6}{Department of Astronomy and Astrophysics,
The Pennsylvania State University, University Park, PA 16802}

% ABSTRACT ---------------------------------------------------------------------
\begin{abstract}
The $z=3.02$ quasar SDSS J095253.83+011421.9 exhibits broad metal-line emission
(\civ\ FWHM$\simeq$9000 \kms), but broad \lya\ emission is not present. Instead,
only a narrow \lya\ line is observed (FWHM$\simeq$1140 \kms).  The large
\civ/\lya\ ratio in the broad-line region (BLR) emission from this object
can be matched most
closely by a BLR dominated by gas at very high densities ($10^{15}$\,cm$^{-3}$),
which suppresses the \lya\ emission, and illuminated by an incident power-law
extending to $\sim$200\,\micron, which yields increased emission from purely
collisionally excited coolant lines (such as \civ, \Nv\ and \ovi) but not from
recombination lines like \lya.  However, the strong \Ciii\ emission predicted
by this model is not observed, and the observed broad \ciii\ emission must come
from a lower-density BLR component and should be accompanied by broad \lya\ 
emission which is not observed.  The least unlikely explanation for this 
spectrum seems to be that any intrinsic broad \lya\ emission is removed by 
smooth \Nv\ absorption in the red wing of the \lya\ emission line and by smooth
\lya\ absorption in the blue wing of the \lya\ emission line.  This postulated
smooth absorption would be in addition to the strong, associated, narrow 
absorption seen in numerous ions.  Smooth absorption in \lya, \Nv\ and \ovi\ 
but not in \civ\ would be unusual, but not impossible, although it is suspicious
that the postulated absorption must almost exactly cancel the postulated
intrinsic broad emission.  We conclude that the spectrum of \nlya\ appears 
unique (among $\simeq$3600 SDSS spectra of quasars at $z>2.12$) because of some 
{\em combination} of unusual parameters, and we discuss possible observations 
to determine the combination of circumstances responsible for the spectrum.
\end{abstract}
 %SDSS J134808.79+003723.2, 

% INTRODUCTION -----------------------------------------------------------------
\section{Introduction}  \label{INTRO}

One of the goals of the Sloan Digital Sky Survey 
\markcite{yor00}(SDSS; {York} {et~al.} 2000)
is to obtain spectra for $\sim$10$^5$ quasars, in addition to the $\sim10^6$
galaxies which comprise the bulk of the spectroscopic targets.  From
astrometrically calibrated drift-scanned imaging data
\markcite{gun98,sdss153}({Gunn} {et~al.} 1998; {Pier} {et~al.} 2003) on the
SDSS $ugriz$ AB asinh magnitude system \markcite{fuk96,sdss26,sdss82,sdss85,sdss105}({Fukugita} {et~al.} 1996; {Lupton}, {Gunn}, \& {Szalay} 1999; {Hogg} {et~al.} 2001; {Stoughton} {et~al.} 2002; {Smith} {et~al.} 2002),
quasar candidates are selected primarily using color criteria designed to
target objects whose broad-band colors are different from those of normal
stars and galaxies \markcite{sdssqtarget}({Richards} {et~al.} 2002).  

The First Data Release of the SDSS (DR1; \markcite{dr1}{Abazajian} {et~al.} 
2003) includes fluxed, wavelength-calibrated spectra of $\sim$17,000
quasars \markcite{dr1q}({Schneider} {et~al.} 2003).  This enormous sample 
includes some quasars with unusual
properties. One example is SDSS J095253.83+011421.9 (hereafter SDSS J0952+0114),
whose spectrum shows both broad and narrow metal-line emission but only narrow
\lya\ emission (Fig. \ref{f_spec}).  In this paper we
investigate possible explanations for the missing broad \lya\ in \nlya, namely:
dust in the broad-line gas (\S\,\ref{DUST}),
anisotropic \lya\ emission (\S\,\ref{RAD}),
broad-line gas with unusual physical properties such
that \lya\ is intrinsically weak (\S\,\ref{PAR}), 
an unusual spectrum incident on the broad-line region (\S\,\ref{IR}), or
an absorption effect wherein \Nv\ and \lya\ absorption
remove the red and blue wings of broad \lya, respectively (\S\,\ref{ABS}).
We summarize our conclusions in \S\,\ref{CON}.

\section{Observations and Basic Parameters}	\label{SPEC}

Relevant instrumental and observational details of the SDSS can be found in
\markcite{dr1q}{Schneider} {et~al.} (2003). Here we simply summarize the
results on \nlya, which was targeted for spectroscopy as a $z\gtrsim3$
quasar candidate.  It was undetected
in the FIRST survey \markcite{bwh95}({Becker}, {White}, \& {Helfand} 1995),
and with a magnitude of $i=18.95$
\markcite{dr1q}({Schneider} {et~al.} 2003)
it is radio-quiet, with radio-loudness parameter $R_i<1.024$
\markcite{sdss1st}({Ivezi{\' c}} {et~al.} 2002).

\begin{figure*}
\plotone{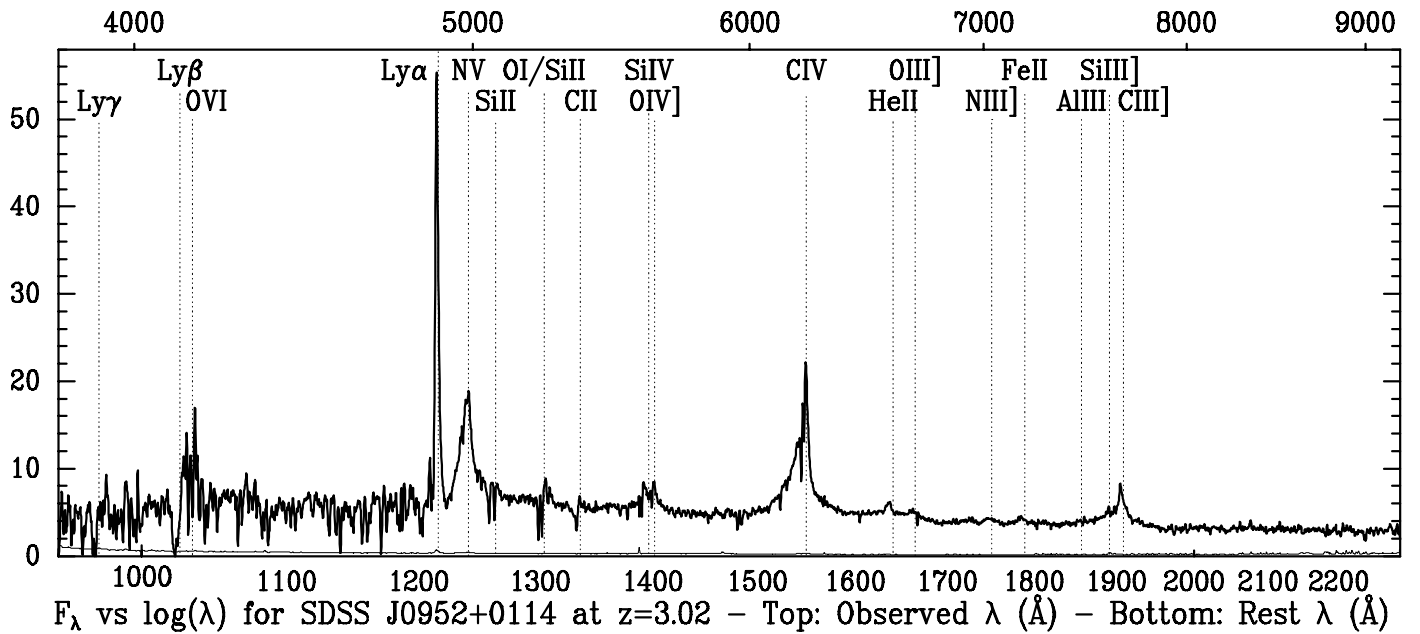}
\caption{Coadded SDSS spectrum of SDSS J095253.83+011421.9 (thick line)
and error array (the barely visible thin line at the bottom of the plot),
both smoothed by a 3-pixel boxcar filter.
\lya\ is resolved at the spectral resolution of $R\simeq2000$.
On this logarithmic wavelength scale, 
emission lines of equal velocity widths have the same plotted widths.
Rest-frame wavelengths in Angstroms %\AA\ 
are shown along the bottom axis, and observed-frame wavelengths along the 
top axis.  The ordinate $F_{\lambda}$ has units of 
10$^{-17}$ erg s$^{-1}$ cm$^{-2}$ \AA$^{-1}$.}\label{f_spec}
\end{figure*}

\begin{figure*}
\plotone{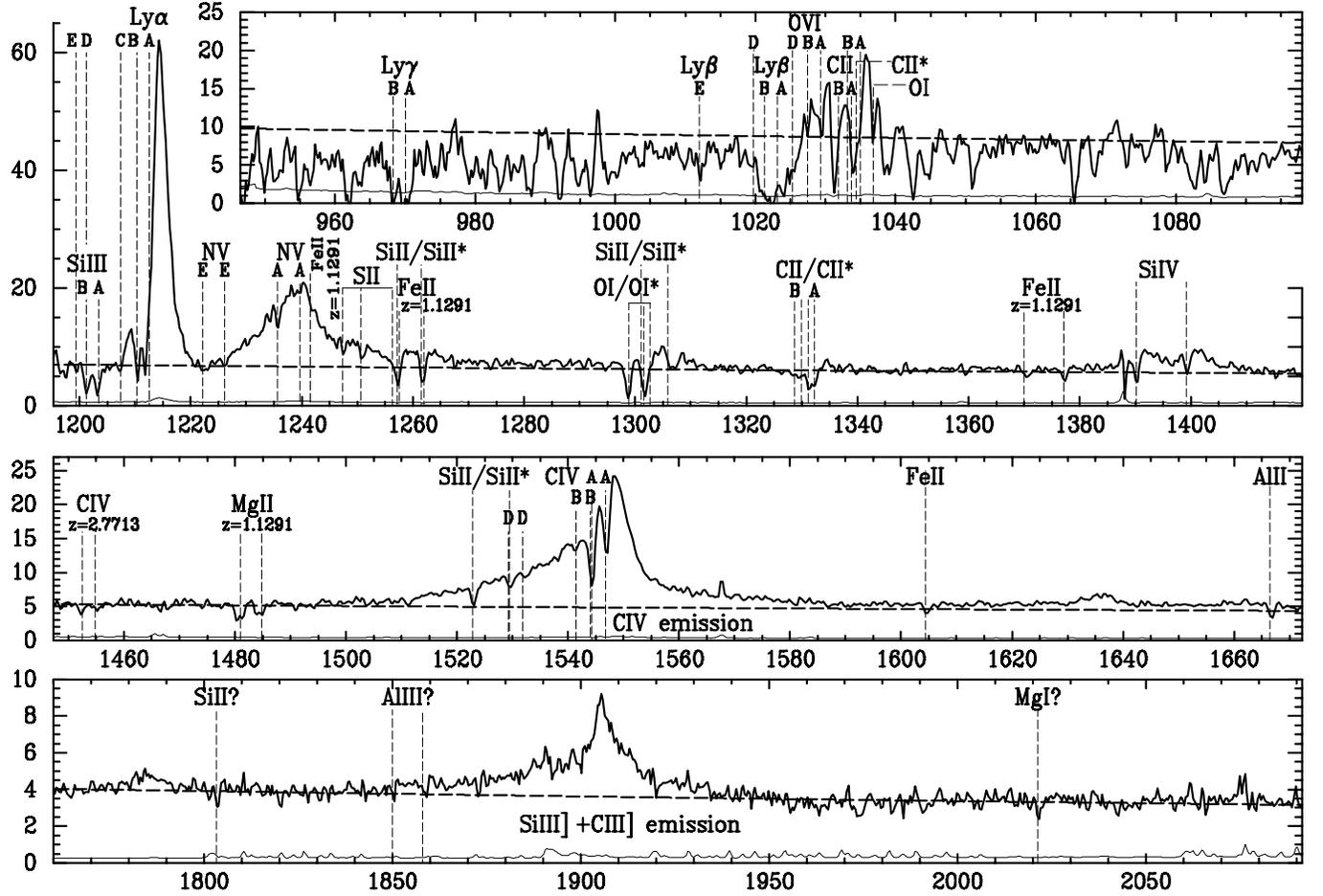}
\caption{Detailed breakdown of the absorption spectrum of \nlya.  
The unsmoothed spectrum and its error array are plotted as the thick and thin
lines, respectively, over four panels which are slightly noncontiguous in
wavelength.  Our power-law continuum fit is plotted as the dashed line.
The labels ABCDE denote the different associated absorbers (see Table
\ref{t_abs}); unlabeled absorption lines are from system A.
Intervening absorption is marked with the redshift of absorption.
The abscissas are rest wavelengths at the systemic $z=3.02$.
The ordinates are $F_{\lambda}$ in units of
10$^{-17}$ erg s$^{-1}$ cm$^{-2}$ \AA$^{-1}$.}\label{f_split}
\end{figure*}

Spectra were obtained twice using the SDSS spectrographs, on Modified Julian
Dates 51608 and 51908, in very similar seeing conditions.  The latest
spectrophotometrically calibrated versions of these 
spectra\footnote{The earlier versions of the spectra published in the SDSS DR1
(http://www.sdss.org/dr1/) show a $\simeq$15\% flux difference due to
spectrophotometric calibration errors 
\markcite{dr2}({Abazajian} {et~al.} 2004).} differ in that the continuum of the
earlier spectrum is significantly bluer (see below).
For our analysis, we coadded the latest versions of the two SDSS spectra
of \nlya, weighting by the inverse variance at each pixel (Fig. \ref{f_spec}).
We adopted a systemic redshift of $z = 3.020 \pm 0.005$ from an initial fit to
the narrow components of \lya, \Nv, and \ciii.  This redshift is statistically
identical to the $z=3.022$ given in \markcite{dr1q}{Schneider} {et~al.} (2003).
We fit a power law from 1348\,\AA\ to 2257\,\AA\ rest frame, excluding broad 
emission line wavelength regions listed in Table 2 of Vanden Berk et al. (2001),
hereafter V01.  We found a best-fit power-law spectral index 
($F_{\lambda} \propto \lambda^{\alpha_{\lambda}}$)
of $\alpha_{\lambda}=-1.32\pm0.23$.
This uncertainty is conservative in the sense that it assumes that 
the different exponents for power-law fits to the individual SDSS spectra 
($\alpha_{\lambda}=-1.45\pm0.03$ and $\alpha_{\lambda}=-1.13\pm0.03$, 
respectively)  are caused entirely by spectrophotometric errors, 
when in fact they could be caused entirely by variability.
Between \lya\ and \Nv\ there is a window $\sim$1350\,\kms\ wide where the
observed spectrum matches this power-law continuum fit (see Fig. \ref{f_split}),
which indicates that the \lya\ and \Nv\ emission lines are not significantly
blended. We normalized the spectrum by our power-law fit to the continuum before
analyzing the emission and absorption lines in detail (see \S\,\ref{FIT}).

Figure\,\ref{f_spec} shows that \nlya\ has a narrow emission-line component
(seen in \lya, \civ, and other metal lines) plus an asymmetric, broad
emission-line component seen in \Nv, \civ\ and \ciii\ but not in \lya.
Emission lines in individual quasars have similar but not necessarily
identical profiles.  Just how much of an outlier \nlya\ is in the distribution
of relative linewidths would require a detailed study of the SDSS
quasar sample beyond the scope of this paper.
In any case, no other similarly extreme spectrum exists among the $\sim$3600
quasars in the SDSS Data Release Two for which \lya\ falls within the spectral
coverage of the SDSS spectrographs ($z\geq2.12$).

\section{Possible Models for \nlya}	\label{WASSUP}

\subsection{Dust in the Broad-Line Region (BLR)?}	\label{DUST}

If the gas responsible for the broad components of the emission lines contains
dust, and has a sufficient optical depth to \lya\ at all velocities,
the broad-component \lya\ photons could be destroyed by the dust
before escaping.  (\lya\ is a resonance line ---
a strong, permitted transition from the ground state of an ion ---
and thus has a small mean free path for absorption or scattering.)
This effect cannot be ruled out, but it appears unlikely for several reasons.
Most dust cannot survive long in typical high-density, high-ionization
BLR conditions \markcite{ld93}({Laor} \& {Draine} 1993) and would
be more likely to be found in the narrow-component gas, given the trend for
narrower lines to arise farther from the central engine
\markcite{pw00}({Peterson} \& {Wandel} 2000).
There may be rare cases of objects with transient dust in their BLR,
since dust destruction is not instantaneous.
However, \NV\ and \CIV\ are also resonance lines that should
be affected by dust.  That is, dust by itself cannot
make the \civ/\lya\ or \Nv/\lya\ ratios arbitrarily large (at least not without
reducing all the lines' equivalent widths far below their observed values).
Narrow-line AGNs with weak \lya\ attributed to resonant scattering
and destruction by dust, rather than just to \lya\ absorption,
have $1\lesssim$\,\civ/\lya\,$\lesssim$1.4
\markcite{dsd95,deb01,dea03}({Dey}, {Spinrad}, \& {Dickinson} 1995; {De Breuck} {et~al.} 2001; {Dawson} {et~al.} 2003).
This is much smaller than the \civ/\lya\,$\simeq$\,$15\pm5$ measured for
the blueshifted broad-component gas in \nlya, at $2000<v<5000$\,\kms.
Therefore, while dust could contribute to the weakness of the
broad \lya\ in \nlya, it is not likely to be the dominant effect.

\subsection{Anisotropic Emission?}	\label{RAD}

If the broad-component gas contains a hydrogen ionization front, the \lya\ will
be emitted only from the illuminated face of the cloud (with source function
$S_{\nu} \propto \cos \theta$), while \civ\ can be emitted more isotropically.
For example, with $N_H=10^{23}$\,cm$^{-2}$, log\,$U\simeq0$ ($U$ is the ratio
of ionizing photon to total hydrogen densities), and densities
$n_H\lesssim10^{9.5}$\,cm$^{-3}$, \civ\ is nearly isotropic and will have a 
rest-frame equivalent width approximately equal to that observed in \nlya\ (see
\S\,2.6 and Fig. 7 of \markcite{kea97}{Korista} {et~al.} 1997a).
Given such conditions, it may be possible
to construct kinematic BLR models such that from some viewing directions the
\lya\ profile is dramatically narrower than the \civ\ profile.

A full study of this possibility would require
coupling photoionization studies incorporating anisotropic emission
with specific geometric and kinematical models of the BLR. 
However, after consideration of numerous toy models, we do not believe that the
spectrum of \nlya\ can be reproduced by recourse solely to anisotropic emission.
The most anisotropic toy model is one where the BLR is a wind
flowing off an optically thick accretion disk at a 45\arcdeg\ angle with 
velocity $v$ and is viewed at that same angle above the disk.
Because of the anisotropy of \lya, the part of the wind nearest to us
(azimuthal angle $\phi=0\arcdeg$)
does not emit \lya\ along our sightline, only
isotropic \civ\ (blueshifted by velocity $v$).
The part of the wind farthest from us ($\phi=180\arcdeg$) emits \lya\ as well
as \civ\ along our sightline, with zero projected velocity and only turbulent
broadening, since our sightline is normal to the outflow direction there.
Summed over all $\phi$, this model produces blueshifted \civ\ emission which is
broader than \lya. However, it cannot produce the red wing of the \civ\ emission
line, which is redshifted relative to \lya.  Nor can it produce \lya\ narrow
enough to match the observations: the maximum observed \civ\ emission velocity
($\sim$9000\,\kms) corresponds to $v$ in this model,
and at a line of sight velocity of $\frac{1}{2} v$ the \lya\ emission
is down from its peak strength by only a factor of $\simeq2.8$,
compared to a factor of $\gtrsim50$ in the observed spectrum.
Similarly, a central obscuring disk embedded in an outflowing,
ionization-stratified BLR can produce a blue wing of \civ\ without strong 
\lya\ (Fig. 26 of \markcite{agn90}{Netzer} 1990) --- essentially by making the
emission anisotropic after the fact --- but it cannot produce a broad red wing
of \civ\ without \lya, nor can it explain linewidth differences as large as
seen in \nlya.  Thus, anisotropic \lya\ emission alone is unlikely to explain 
the unusual spectrum of \nlya.

\subsection{Unusual Physical Conditions in the Broad Line Region?}  \label{PAR}

\markcite{kea97}{Korista} {et~al.} (1997a) have used the photoionization code
CLOUDY \markcite{c90}({Ferland} {et~al.} 1998) to compile
plots of predicted quasar BLR emission-line rest-frame
equivalent widths (REWs) as functions of the column density, incident 
ionizing spectrum, and metal abundance of the emitting gas clouds.
We used these plots to investigate whether significant \Nv, \civ, and \ciii\ 
emission without accompanying \lya\ emission is expected to occur
anywhere in the range of parameter space occupied by typical quasar BLRs.
We were unable to find any regions of parameter space which simultaneously met
the observed constraints on the total REW$_{C\,IV}$, REW$_{Ly\,\alpha}$ 
and the line ratio \civ/\lya.

It is interesting to compare these results with detailed modelling of the
quasar Q~0207$-$398 by \markcite{fea96}{Ferland} {et~al.} (1996).  
Q~0207$-$398 has a very broad emission component with \civ/\lya\,$\simeq3$
\markcite{bea96}({Baldwin} {et~al.} 1996).  The CLOUDY simulations of
\markcite{fea96}{Ferland} {et~al.} (1996) and our own CLOUDY simulations show
that a \civ/\lya\ ratio of $\simeq3$ is the maximum which can be produced over
a wide range of $N_H$ for metallicities between half and five times solar.  
Thus, the measured \civ/\lya\,$\simeq15\pm5$ in the broad emission component
of \nlya\ cannot be explained in the same manner as the
\civ/\lya\,$\simeq3$ of Q~0207$-$398.
Dust destruction of \lya\ will not substantially increase the
maximum \civ/\lya\ ratio because the maximum ratio only occurs for optically
thin clouds in which hydrogen is fully ionized and \lya\ photons are not
resonantly trapped.  \lya\ photons are therefore more likely to escape 
from such clouds than to be destroyed by dust.

However, the above results come from photoionization simulations which extend
only to densities $n_H=10^{14}$ cm$^{-3}$ and column densities 
$N_H=10^{24}$ cm$^{-2}$, and different behavior may be seen when both parameters
are an order of magnitude or more larger. \lya/\hb\ ratios $\sim$100 times lower
than the typical observed value of $\sim$10 \markcite{fer99}({Ferland} 1999)
can be produced in photoionization simulations of accretion disks 
reaching $N_{H}=10^{25}$ cm$^{-2}$ and $n_H=10^{15}$ cm$^{-3}$ 
\markcite{rbcs92}({Rokaki}, {Boisson}, \&  {Collin-Souffrin} 1992).
\lya\ photons are trapped by resonant scattering at such high column densities, 
and as the density increases above $n_H=10^{11}$ cm$^{-3}$ the \lya\ REW 
steadily decreases due to increasing collisional deexcitation of the $n=2$
level of \ion{H}{1} and bound-free absorption by neutral metals or
\ion{H}{1} in an $n\geq2$ level (see \S\,3.3 of
\markcite{dcs90iv}{Dumont} \& {Collin-Souffrin} 1990 or Fig. 5 of 
\markcite{fer99}{Ferland} 1999).  Evidence for emission from gas at such high
densities has previously been found by \markcite{arp102b}{Halpern} {et~al.}
(1996) in the broad-line radio galaxy Arp 102B, which shows broad, 
double-peaked emission line components in \hb\ and \civ\ but only 
single-peaked emission in \lya\ (their Fig. 3).

To test the viability of high-$n_H$, high-$N_H$ models for \nlya, we ran 
photoionization simulations with CLOUDY
(using its standard incident AGN spectrum)
for densities up to $10^{16}$\,cm$^{-3}$ and column densities up to
$10^{25}$\,cm$^{-2}$ over a wide range of ionization parameters.
The highest \civ/\lya\ ratio we found for optically thick clouds with \ion{H}{1}
ionization fronts ($N_H\geq10^{23}U$\,cm$^{-2}$) was \civ/\lya$\sim$6,
for $n_H=10^{15}$\,cm$^{-3}$ and $-0.5<\log\,U<0.5$.  (Increasing the column
density above the optically thick limit does not affect \civ/\lya, though it
may affect ratios of low-ionization line strengths relative to \lya.)  However, 
even with 100\% covering of the ionizing source by the BLR, the REW$_{C\,IV}$
for such models is only 0.9--3.8\,\AA.  This is at least an order of magnitude
lower than the broad component REW$_{C\,IV}$=42\,\AA\ observed in \nlya.
		Slightly higher \civ/\lya\ ratios ($\sim$8) can be reached in
		clouds with column densities an order of magnitude below
		the optically thick limit, but the REW discrepancies are even
		worse in such cases.
	The `best fit' modelled and observed numbers could be reconciled only
	if the continuum flux in \nlya\ was suppressed by at least an order of
	magnitude, but there is no evidence that is the case.
We conclude that high densities
alone cannot simultaneously explain the abnormal \civ/\lya\ line ratio
and the normal \civ\ and \lya\ REWs in \nlya.

\subsection{An Unusual Spectrum Incident on the BLR?} \label{IR}

The \civ/\lya\ ratio in the broad-line gas in \nlya\ is a factor of 2.5$\pm$0.8
higher than in our `best fit' photoionization model ($n_H=10^{15}$\,cm$^{-3}$,
$N_H\gtrsim10^{24}$\,cm$^{-2}$ and $-0.5<\log\,U<0.5$),
and the \civ\ REW is at least an order of magnitude higher. 
\markcite{fer99}{Ferland} (1999) discusses the spectrum of an ionized cloud 
as a function of additional physical parameters besides those considered in
\markcite{kea97}{Korista} {et~al.} (1997a).  Unusual values of such additional
parameters might help explain the observations.  Inspection of the various
plots in \markcite{fer99}{Ferland} (1999) shows that the {\em only} parameter
which might substantially affect the observables \civ/\lya\ and REW$_{C\,IV}$
is the assumed long-wavelength cutoff wavelength of the incident power-law
(his Fig. 3).
Extending this cutoff from its arbitrary default value of 0.912\,\micron\ 
to $\geq$200\,\micron\ would increase \civ/\lya\ by a factor of two or more,
and REW$_{C\,IV}$ by a factor of five or more.
The greater infrared flux increases the
temperature of the gas through free-free heating without increasing its
ionization.  For a sufficiently long cutoff wavelength ($\geq$200\,\micron),
the flux from purely collisionally excited coolant lines like \civ, \Nv\ and
\ovi\ increases more than the flux from recombination lines like \lya\ (though
\lya\ can be collisionally excited as well, of course).  
The SED required by this scenario is consistent with the observed range of
quasar SEDs \markcite{elv94}({Elvis} {et~al.} 1994).  The putative far-IR
power-law cannot extend to radio wavelengths, however, since \nlya\ is 
radio-quiet.

The emission-line REWs of \nlya\ are mostly consistent with this scenario.
Compared to the composite SDSS quasar spectrum of
\markcite{sdss73}{Vanden Berk} {et~al.} (2001),
\nlya\ has a total \lya\ REW a factor of three smaller than average and a total
\civ\ REW a factor of three higher.  However, the REW of the broad \ciii\ 
emission is quite high, and such emission cannot arise in high density gas since
the upper state of that transition is collisionally deexcited at densities
$n_e>10^{12}$ cm$^{-3}$ (conversely, \Ciii\,$\lambda$977\,\AA\ is a strong
coolant at high densities but is not observed to be unusually strong here).
Moreover, broad \ciii\ emission should be accompanied by broad \lya\ emission
with at least \lya/\ciii=5, which is inconsistent with observations.

We can think of two explanations for this discrepancy.  The first is a decrease
in the BLR gas density at the \ion{H}{1} ionization front where the \ciii\ 
emissivity is highest, which would maximize the observed ratio of \ciii/\lya. 
Coupled with high densities in the BLR up to the \ion{H}{1} ionization front
%(which would suppress \lya\ emission) 
and an incident power-law extending well into the far-IR,
%(which would increases the \civ/\lya\ ratio), 
this scenario might simultaneously explain the unusually large broad-line
\civ/\lya\ ratio, the low \lya\ REW, and the observed broad \ciii.  However, it
still predicts much stronger \Ciii\,977\,\AA\ emission than observed, and a
model which invokes such a specific and ad hoc combination of parameters cannot
be considered robust, even for a uniquely unusual quasar from a sample of
$\sim$3600.

The second possible explanation for the lack of broad \lya\ accompanying the
observed broad \ciii\ is that the broad \lya\ exists but is hidden by 
absorption, a possibility we consider in detail in the next section.

\subsection{Intrinsic Absorption?}	\label{ABS}

%A last possibility is that \nlya\ has 
\nlya\ may have a \lya\ profile which merely
{\em appears} narrow due to moderately broad, blueshifted absorption 
from \lya\ (in the blue wing of the \lya\ emission line)
and from \Nv\ (in the red wing of the \lya\ emission line).  Such absorption
could originate in the host galaxy or in an outflow related to the quasar.
%(For an example of the latter sort of object, see the broad absorption
%line quasar SDSS J134808.79+003723.2, which is in the sample of
%\markcite{dr1q}{Schneider} {et~al.} (2003) and whose spectrum can be
%accessed via the SDSS website.)  
To investigate this possibility, we modelled the emission and absorption lines
seen in \nlya. %, beginning with the narrow, associated absorption lines.

\subsubsection{Fitting the Spectrum}	\label{FIT}

A single Gaussian was used 
to represent the profile of each absorption line and of most emission lines.
Some emission lines required more than one Gaussian: two
were used for \ciii, and
three for \lya, \Nv, \SIiv\ and \civ\ (three components are often used to model
quasar emission lines; see, e.g., \markcite{bwfs94}{Brotherton} {et~al.} 1994).
Strong emission lines from Table 2 of V01 were included in the fit, along with
a broad Gaussian near 1600\,\AA\ (to fit \feii\ multiplets there) and another
near 1280\,\AA\ (presumably due to \feii\ multiplet UV9).  The parameters of
the emission and absorption line fits are given in Tables \ref{t_em} and
\ref{t_abs}, respectively.\footnote{We {\em tentatively} detect emission from 
\ion{O}{5}]~$\lambda$1218.344 on the red shoulder of \lya, with FWHM=300\,\kms.
Highly ionized gas is expected to have 0.1$\leq$\ion{O}{5}]/\lya$\leq$0.3
\markcite{fea92}({Ferland} {et~al.} 1992) or, equivalently,
\ion{O}{5}]/\ion{N}{5}$\sim$1 \markcite{joe94}({Shields} 1994).
Because \ion{O}{5}] is located only 657\,\kms\ longward
of \lya, previously only limits have
been put on its strength (e.g., \ion{O}{5}]/\lya$\leq$3-10\% for the five
low-redshift quasars studied by \markcite{laor94}{Laor} {et~al.} 1994).
For the narrow emission-line component of \nlya, our value
of \ion{O}{5}]/\lya$\simeq$0.08 is barely consistent with predictions.
For the broad emission-line component, the
\markcite{joe94}{Shields} (1994) prediction of \ion{O}{5}]/\ion{N}{5}$\sim$1 is
ruled out --- we do not detect any broad emission attributable
to \ion{O}{5}] (or \lya).
Either the photoionization models are inaccurate, 
the broad emission-line component in \nlya\ does not
arise in optically thin high-ionization gas, or some mechanism such as
overlying absorption is suppressing broad \ion{O}{5}] along with broad \lya.} 
We identify five blueshifted absorption systems, denoted A--E, in \nlya.
The narrow absorption in the spectrum is shown in detail in
Figure \ref{f_split} and is discussed in detail in Appendix \ref{APP}.

\subsubsection{Line Profiles Corrected for Narrow Absorption} \label{UNABS}

\begin{figure*}	% patnew.sm
%\epsscale{1.15}
\epsscale{1.05}
%\epsscale{1.5}
\plottwo{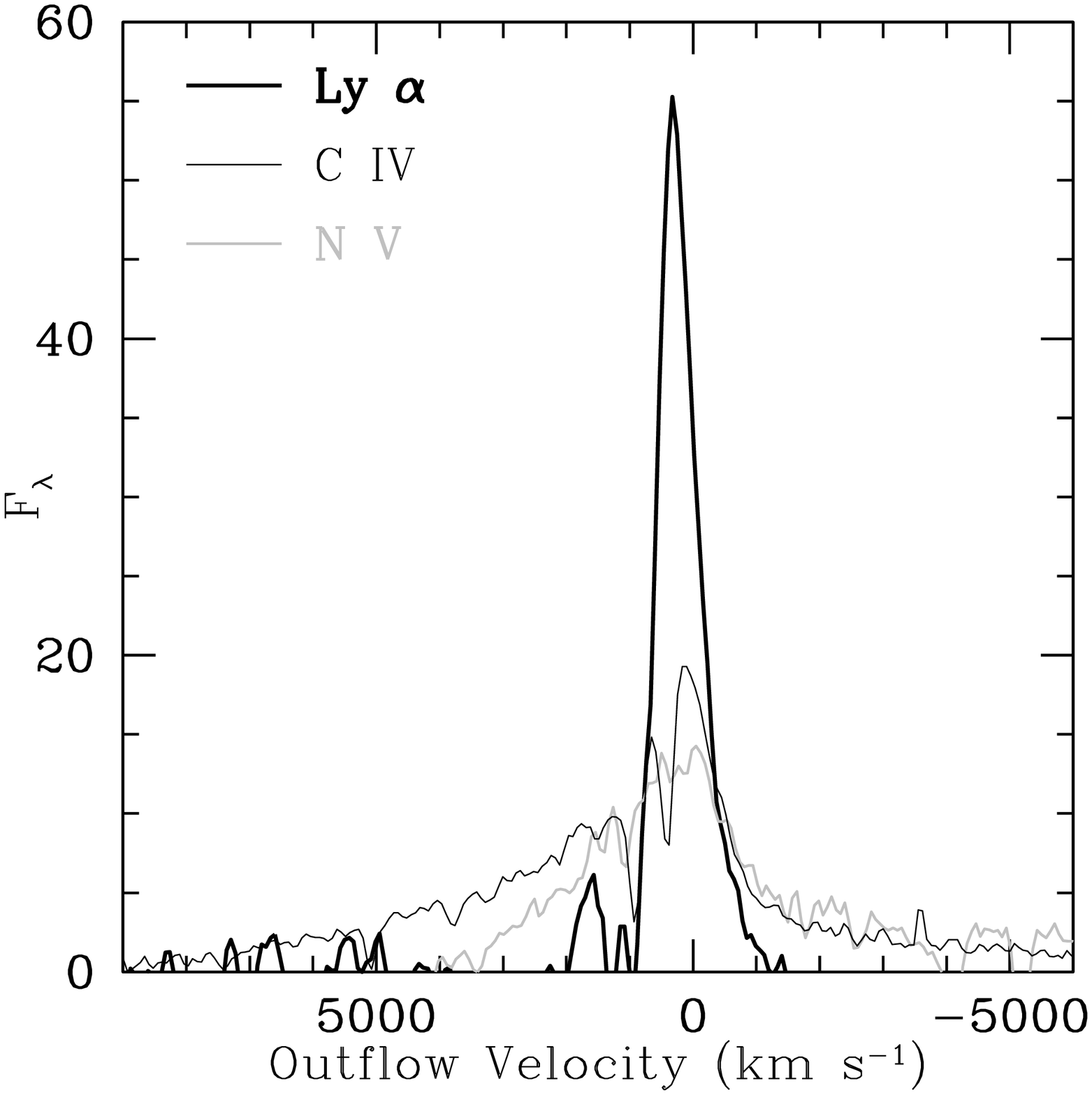}{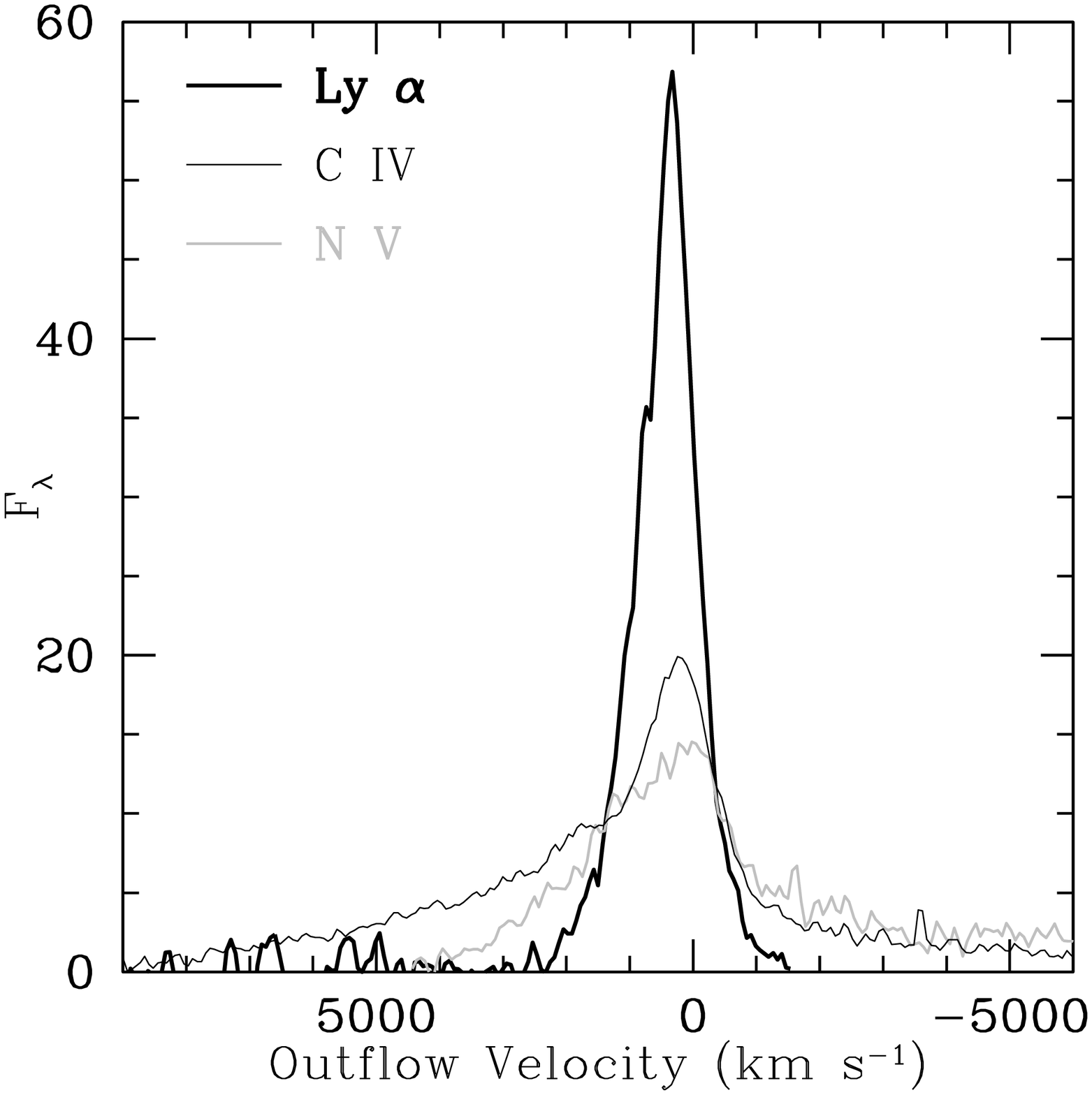}	% b/w
\caption[]{a) \fl\ vs. outflow velocity for the raw, unsmoothed \lya, \civ\ and
\Nv\ emission lines, with $v=0$ set by the laboratory wavelengths of each line
given in Table 2 of \markcite{sdss73}{Vanden Berk} {et~al.} (2001).  To avoid 
repeated plotting of the \lya\ and \Nv\ emission lines, \lya\ is plotted only 
between $-$1600 and 9000 \kms\ and \Nv\ only between $-$5000 and 4450 \kms.
b) Same as a), except using the repaired, unsmoothed spectrum, where all
fitted absorption lines have been removed.
}\label{f_repair}
\end{figure*}

To study the unabsorbed line profiles, we ``repaired'' the spectrum by
reversing the sign of the amplitude of our single-Gaussian fits to the 
absorption lines and adding those Gaussians to the observed spectrum.
The ``repaired'' spectrum is therefore our best estimate of the
spectrum as it would have been seen without narrow absorption.
Figure \ref{f_repair}a plots the \lya, \civ\ and \Nv\ regions of 
the raw spectrum of \nlya\ 
as a function of outflow velocity relative to their
laboratory wavelengths (Table 2 of V01) at $z=3.02$.
The same plot using the ``repaired'' spectrum
is shown in Figure \ref{f_repair}b.
{\em Even after removing the effects of narrow, associated absorption,}
\lya\ is narrower than either \civ\ or \Nv.  
These figures also show that the \civ\ emission line has a strong blue wing,
extending to $\sim$9000\,\kms, which is not seen in \lya\ or \Nv.

\subsubsection{Additional Smooth Absorption?} \label{SMOOTH}

Fitting the emission and narrow absorption lines in \nlya\ with Gaussians has
not revealed an intrinsic \lya\ emission profile as broad as that of \ciii, let
alone \civ.  However, there is some suggestion in Figure \ref{f_repair}b that
there may be additional smooth absorption in \lya\ and \Nv.  The \Nv\ and \civ\ 
emission profiles agree longward of the emission-line peaks, and the \lya\ 
profile has five maxima at $5000<v<9000$\,kms\ which approach the \civ\ profile,
though these maxima may just be noise.
Smooth absorption at $600\lesssim v \lesssim 6500$\,\kms\ in \lya\ and \Nv\ but
not in \civ\ could bring the intrinsic profiles of all three lines into
agreement.   Such absorption is possible: \Nv\ and \ovi\ have very similar
ionization potentials and will be present in the same gas, but \civ\ has
a lower ionization potential and will not necessarily accompany them.

To test this scenario, we assume that there is no smooth absorption in \civ\ and
that our repaired \civ\ emission-line profile represents the intrinsic profile
of all emission lines in \nlya.  We shifted, scaled and added the repaired
\civ\ profile to our power-law continuum fit to the observed spectrum
to best match the observed peaks of \ovi, \Nv\ and \lya\ as follows.
We reconstructed \ovi\ as two doublets, each with 75\% of the total \civ\ flux,
then multiplied the reconstructed spectrum at those wavelengths by 0.75 to
%account for \lya\ forest absorption by matching the surrounding continuum.
match the surrounding continuum and thus account for \lya\ forest absorption.
We reconstructed \Nv\ as two doublets, each wth 45\% of the total \civ\ flux,
and broad \lya\ as a single line with flux equal to that of \civ.

\begin{figure*} % patnew.sm
%\epsscale{1.10}
%\epsscale{1.00}
\epsscale{0.55}
%\plotone{rawlong.eps}\\
%\plotone{replong.eps}\\
%\plotone{repshort.eps}
\plotone{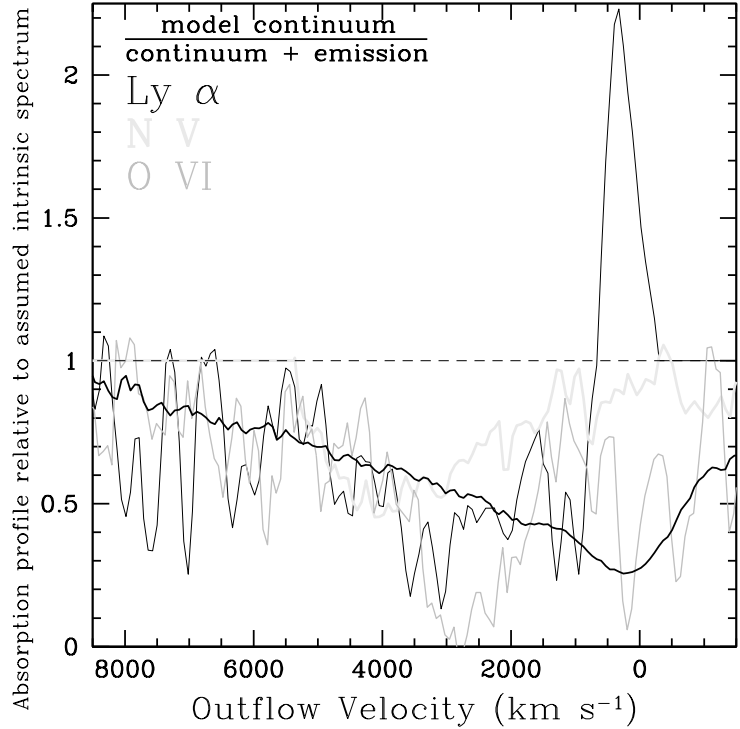}
%\plotone{rawshort.eps}
\caption[]{ %a) 
Estimated {\em total} absorption profiles in \lya\ (dark line),
\Nv\ (light line) and \ovi\ (medium line) as a function of outflow velocity (in
\kms), aligned using the shorter-wavelength member of each doublet line so that
the high-velocity edges of absorption troughs can be compared.
%aligned using the longer-wavelength member of each doublet line so that the
%low-velocity edges of absorption troughs can be compared.
\lya\ and \Nv\ profiles have not been plotted at velocities where they are 
confused.
%b) Estimated {\em smooth} absorption profiles for the same three lines.
%c) Estimated {\em smooth} absorption profiles 
%aligned using the shorter-wavelength member of each doublet line so that the 
%high-velocity edges of absorption troughs can be compared.
The darkest plotted %dash-dotted
line shows the absorption profile needed to completely remove
the repaired \civ\ emission-line profile; i.e., it is our power-law continuum
fit divided by that continuum fit plus the repaired \civ\ emission line profile.
}\label{f_rawshort}
\end{figure*}

We then divided the observed spectrum by this assumed intrinsic spectrum.
The result is an estimate of the total absorption profile required to explain
the observed line profiles of \lya, \Nv\ and \ovi\ as a function of velocity.
Figure \ref{f_rawshort}
shows these total absorption profiles, aligned using the
shorter-wavelength member of each doublet line so that the high-velocity edges
of absorption features can be compared in different lines.
As suggested above, \lya\ at $v>5500$\,\kms\ is
consistent with a recovery to an intrinsic profile matching \civ\ (the 
horizontal line), punctuated by narrow \lya\ forest absorption.
In the outflow velocity region $3500<v<5500$\,\kms, we have reliable
information on all three lines.  (At higher velocities \Nv\ is confused with
\lya, and at lower velocities \ovi\ is confused with a \lyb\ trough which
reaches zero flux; see Fig. \ref{f_split} and Appendix \ref{APP}.)
At $3500<v<5500$\,\kms, all three absorption profiles are consistent with the
darkest line, which is the absorption profile needed to completely remove
the repaired \civ\ emission-line profile.  In other words, %there is no evidence
%for emission at $3500<v<5500$\,\kms\ in \ovi, \lya\ or \Nv.  Therefore,
if we assume the intrinsic emission profile in \ovi, \lya\ and \Nv\ equals
that of \civ, to match the observations in those lines requires absorption
which exactly cancels the assumed emission at velocities $3500<v<5500$\,\kms\ 
(and also at velocities $v>5500$\,\kms\ in \ovi).

Broad absorption line troughs have never been seen to unambiguously cover only
the BLR and not the continuum source, which is understandable since the BLR is
thought to surround the much smaller continuum source.  Thus, if smooth
absorption is really present in \lya\ and \Nv\ but not in \civ, which is already
unusual, the absorption profiles must also almost exactly cancel out the
emission profiles, by remarkable coincidence.  %However, postulating such
%properties may not be so improbable when it explains the unusual properties
%which initially drew our attention to this one quasar among $\sim$3600.

Additionally, if smooth absorption is present, it appears to have different
profiles in \lya\ and \Nv, with different covering factors (assuming the
absorption is saturated) at velocities $3000<v<1500$\,\kms, at least.  
This would not be unusual, but is only a tentative result.  The absorption
profiles are systematically uncertain because of the assumption that the
intrinsic emission-line profiles are identical to the repaired \civ\ profile.

In summary, the evidence for additional smooth absorption in \nlya\ is mixed.
Gas outflowing at $\sim$600-5500\,\kms\ absorbing in \ovi, \Nv\ and \lya\ (but
not in \civ, which is unusual) with
peak fractional absorptions of $\sim$45\% {\em can} reconcile those lines'
observed emission-line profiles with intrinsic profiles matching the repaired
profile of \civ.  However, the smooth, strong absorption would have to almost
exactly cancel the broad emission, at least at velocities $3500<v<5500$\,\kms.
The arguments for smooth absorption are that
the red wing of \Nv\ matches the red wing of \civ, 
\lya\ may recover to the \civ\ profile in its far blue wing, and
the lack of broad \ion{O}{5}] and broad \lya\ which should accompany broad \Nv\ 
and broad \ciii, respectively, is difficult to explain except via absorption.

%First, to explain the deficit in the blue wing of \lya\ (compared to the SDSS
%composite quasar) would require a strong \lya\ absorption trough (REW=5.1\,\AA,
%stronger than all but one of the detected absorbers) with an unusually broad
%and shallow shape (FWHM$\sim$1500\,\kms\ and 25\% partial covering).
%Second, there is no sign of metal-line absorption accompanying such a trough.  
%Third, the lack of a broad, red wing of \lya\ emission would require
%additional \lya\ absorption from redshifted gas, as well as blueshifted.
%The missing red wing of \lya\ cannot be due to absorption in \Nv\ because there
%is no sign of the broad \ovi\ absorption which such gas would produce even if
%it was so highly ionized that it produced negligible \civ\ or \lya\ absorption.
%Thus, the absence of broad \lya\ in \nlya\ is not due to
%\lya\ and \Nv\ absorption in the blue and red wings of \lya, respectively.

\section{Conclusions}	\label{CON}

We have discussed the unusual spectrum of \nlya, which
exhibits broad metal-line emission but only narrow \lya\ emission.
We argue that this unusual spectrum cannot be {\em solely} due 
to dust extinction in the BLR,
to anisotropic emission of \lya, 
or to unusual physical conditions in the BLR.
Most but not all
of the spectrum's properties are explainable as emission from a BLR of
predominantly high density gas ($n_H\sim10^{15}$ cm$^{-3}$), which suppresses
\lya, illuminated by an incident power-law continuum extending to
$\geq$200\,\micron, which increases the collisionally excited metal-line
emission.  %This would be one of the highest densities yet inferred for BLR gas
%(e.g., \markcite{oea04}{Ogle} {et~al.} 2004), 
However, the BLR in \nlya\ cannot consist exclusively of high density gas
because the observed broad \ciii\ emission would be collisionally deexcited at
densities higher than $10^{12}$ cm$^{-3}$.  

Although it is not entirely satisfactory, the most plausible explanation we
have found for the apparent lack of broad \lya\ is that it is due to smooth
absorption by \Nv\ in the red wing of \lya\ and by \lya\ in the blue wing of
\lya.  Such absorption must be in addition to the complex, narrow intrinsic
absorption system seen, would have to almost exactly cancel the intrinsic broad
emission (which might mean that it is the first known example of a broad
absorption line trough which covers the BLR but not the continuum source), 
and must be present in \ovi, \Nv\ and \lya\ but not in \SIiv\ or in \civ,
which is quite unusual.  In any case,
it seems that some {\em combination} of unusual parameters is required
to explain \nlya, which helps explain why its spectrum is one of a kind.
%although a comprehensive study of the distribution of line profile differences
%in SDSS quasars is needed to determine just how much of an outlier \nlya\ is.

The X-ray spectrum of \nlya\ should consist of a typical power-law plus
absorption if the postulated highly-ionized, smooth absorption is real.
On the other hand, if its line ratios are largely intrinsic and the trends of
\civ/\lya\ with other observables found by \markcite{bev99}{Wills} {et~al.} 
(1999) can be extrapolated over an order of magnitude to apply to \nlya,
it should have a very soft X-ray spectrum, a very broad H$\beta$ line, weak
optical \feii\ emission, and strong \oiii\ emission.  
\markcite{bev99}{Wills} {et~al.} (1999) suggest all these trends may derive from
a small Eddington parameter (the accretion rate relative to the Eddington rate).
%That scenario seems an unlikely one for \nlya\ if the BLR gas reaches the very
%high densities we infer as a possible explanation.  On the other hand, like
%\nlya, some double-peaked emission-line AGN have very large broad-line
%\civ/\lya\ ratios (\markcite{arp102b,eea04}{Halpern} {et~al.} 1996; Eracleous
%{et~al.} 2004), and they have been argued to be
%low-Eddington-parameter objects \markcite{era04}({Eracleous} 2004), so perhaps
%there is some connection between high BLR densities and low accretion rates.
%%
%%Nonetheless,
%%it may be more plausible that \nlya\ simply represents a BLR dominated by the
%%high-density tail of the distribution of quasar BLR gas densities
%%{\em and} the long-wavelength tail of the distribution of cutoff wavelengths
%%for the power-law continuum illuminating the BLR.
While \nlya\ may not provide direct insight on the typical quasar BLR if it is
indeed a high-density BLR illuminated by an unusual SED, it does help delineate
the range of physical parameter space which BLRs occupy and which must therefore
be incorporated into BLR models.  For example, quasar broad-line \civ/\lya\ 
ratios may be strongly affected by free-free heating of the BLR, but as
\markcite{fer99}{Ferland} (1999) point out, very little work has been done to
investigate the effects of free-free heating on quasar spectra,
despite the fact that it can have a
greater effect on the observed spectrum than the incident X-ray continuum.

The most useful future observation of \nlya\ would probably be flux-calibrated
spectroscopy at higher resolution to better determine the true profiles of
the emission lines, and thus the presence or absence of smooth absorption,
by reducing the confusing effects of the many narrow absorption lines present
(and also to enable physical modeling of the narrow, associated absorbers).
It would also be valuable to obtain an X-ray hardness ratio measurement
and near-IR spectroscopy of \mgii, H$\gamma$ and H$\beta$.  And if \nlya\ has
a BLR exposed to a power-law continuum extending from X-ray wavelengths to
$\geq$200\,\micron, or $\geq$800\,\micron\ observed, SCUBA or {\em Spitzer}
photometry at those wavelengths could search for that continuum directly (with
the caveat that there is good evidence that the continuum illuminating the BLR
in quasars can be different from the continuum seen along our lines of sight to
individual quasars; \markcite{kfb97}{Korista}, {Ferland}, \&  {Baldwin} 1997b).

\acknowledgements
We thank G. Richards, M. Eracleous, J. Charlton and the referee
for helpful discussions.
Funding for the creation and distribution of the SDSS Archive
(http://www.sdss.org/) has been provided
by the Alfred P. Sloan Foundation, the Participating Institutions, NASA,
the National Science Foundation, the U.S.
Department of Energy, the Japanese Monbukagakusho, and the Max Planck Society.
The SDSS is managed by the
Astrophysical Research Consortium (ARC) for the Participating Institutions:
The University of Chicago, Fermilab, the
Institute for Advanced Study, the Japan Participation Group, The Johns Hopkins
University, Los Alamos National Laboratory, the Max-Planck-Institute for
Astronomy (MPIA), the Max-Planck-Institute for Astrophysics (MPA),
New Mexico State University, University of Pittsburgh, Princeton University, the
United States Naval Observatory, and the University of Washington.  P. B. H. was
supported by Fundaci\'{o}n Andes and the Department of Astrophysical Sciences at
Princeton University, M. N. O. by an REU supplement to NSF grant AST-0071091 to
Princeton University, D. J. E. by NSF grant AST-0098577 and an Alfred P. Sloan
Research Fellowship, M. A. S. by NSF grant AST-0307409, and D. P. S. by NSF
grant AST-0307582.

\begin{appendix}
%\appendix
\section{Details of the Narrow Associated Absorbers} \label{APP}

We identify five associated (and two intervening)
narrow absorption systems in \nlya\ 
and denote them A through E in order of increasing outflow velocity.
A higher-resolution spectrum with a high signal-to-noise ratio is needed
to study these absorbers in detail and eliminate the numerous ambiguities
due to blending.  The absorption around the \ovi\ emission line is particularly
complex, and suffers from \lya\ forest contamination as well.
The strongest associated absorption system (A)
includes rare, excited, metastable \oi*, \cii* and \SIii* lines 
(though the putative \SIii*\,$\lambda$1533 seems anomalously weak)
and is blueshifted by $v_A=$770$\pm$100\,\kms\ from the systemic redshift,
neglecting the global uncertainty in the latter.
System B at $v_B=$1310$\pm$10\,\kms\ (which also includes \cii* absorption)
exhibits line-locking with system A \markcite{sca73,bm89}({Scargle} 1973;
{Braun} \& {Milgrom} 1989), since the velocity difference
$\Delta v_{AB}=540\pm100$\,\kms\ 
is consistent with the velocity separation of the \civ\ doublet (497\,\kms).
There is also \lya\ absorption at $v_C=2050\pm190$\,\kms. 
There may be \civ\ absorption at $v_D=3570\pm20$\,\kms, possibly accompanied by
\lya\ (the putative \civ\,$\lambda$1548 is blended with \SIii*\,$\lambda$1533
from system A, and the putative \lya\ 
with \SIiii\,$\lambda$1206.5
absorption from system B).  Finally, there is \lyb, \lya, \Nv\ and possibly
\ovi\ absorption at $v_E=4050\pm40$\,\kms.

One complication with modelling the absorption systems is that at least some 
of them exhibit full covering of the continuum source but only partial covering
of the emission-line region, a common feature in intrinsic absorption systems
(e.g., \markcite{hbjb97}{Hamann} {et~al.} 1997).  At the wavelength of the 
blended \lyb\ absorption
from systems A and B ($z_{blend}=3.0056$; see Table \ref{t_abs}), 
the residual flux over 2-3 resolution elements is consistent with zero
(measuring 2.6$\pm$5.6\% of the normalized intrinsic spectrum),
and similarly at \lyg. 
However, the residual flux
in \lya\ at the same redshift measures much more than 2.6\% %$\pm$5.6\% 
of the estimated intrinsic spectrum at that wavelength.  
This discrepancy can be explained if the
absorption covers the continuum source but not the \lya\ emission-line region.
These different covering factors do not greatly 
affect the fitting but might affect the interpretation of 
absorption doublet line ratios near emission lines.

The physical conditions in the various absorbers may be quite different.
We derive an approximate $N_{H I}$ value for each absorber by finding the
$N_{H I}$ value which produces a Voigt absorption profile 
whose FWHM equals the resolution-corrected FWHM of our fit
to the system's \lya\ absorption (see, e.g., Fig. 14 of \markcite{mhea03}{Mas-Hesse} {et~al.} 2003).
System A has log($N_{H I}$)$\sim$18.7 cm$^{-2}$, absorption from many
low-ionization species (sufficiently strong to suggest solar metallicity), 
relatively weak \ovi\ and \Nv\ absorption, and excited-state absorption
indicating densities of at least $n_e \simeq 10^{2.5}$ cm$^{-3}$.
It may also contain dust: if the tentative
2026\,\AA\ 
absorption is real and
is identified as \znii\ instead of \mgi, there may be accompanying
\znii\,$\lambda$2062, but there is no sign of \crii\,$\lambda$2056 or
\crii\,$\lambda$2066.  If confirmed by a better spectrum, this would
indicate that \crii\ is depleted onto dust in the absorber.
System B is detected in a range of ionization states from \cii\ to \ovi,
but there is no unambiguous evidence for excited-state absorption in it.
System C is
unambiguously detected only in \lya, probably due to a low column density.
Systems D and E appear more highly ionized on average, with possible detections
of \ovi\ and \civ\ in the former and \Nv\ in the latter.

\end{appendix}

% REFERENCES -------------------------------------------------------------------

%\clearpage

%%FIGURES / FIGURE CAPTIONS ----------------------------------------------------

%\clearpage

%%TABLES -----------------------------------------------------------------------
\begin{deluxetable}{llccrcl}
\tablecaption{SDSS J0952+0114 Line Identifications\label{t_em}}
%\tabletypesize{\footnotesize}
\tabletypesize{\scriptsize}
\tablewidth{450.00000pt}
\tablehead{
\colhead{Line}& \colhead{$\lambda_{rest}$\tablenotemark{a}}& \colhead{Redshift}&
\colhead{FWHM\tablenotemark{b}}& \colhead{REW\tablenotemark{c}}
&\colhead{Flux\tablenotemark{d}} &\colhead{Notes\tablenotemark{e}}
}
\startdata
\multicolumn{6}{l}{Emission Line Components}\\
\tableline
C\,III & \phm{1}977.02 & 3.0205$\pm$0.0007 & \phm{~~}470$\pm$90\phm{~~} & $-$0.84$\pm$0.19 & \phm{~~}8$\pm$2 & \\ % 9.35
N\,III & \phm{1}989.790 & 3.0205$\pm$0.0008 & \phm{~~~}540$\pm$110\phm{~} & $-$0.98$\pm$0.26 & \phm{~~}9$\pm$2 & \\ % 9.18
O\,VI & 1033.03 & 3.0215$\pm$0.0006 & \phm{~}2180$\pm$80\phm{~~} & $-$15.88$\pm$0.88 & 138$\pm$8 & \\ % 8.636
Ly$\alpha$ & 1215.6701 & 3.0108$\pm$0.0004 & \phm{~}1950$\pm$30\phm{~~} & $-$7.64$\pm$1.25 & \phm{~}52$\pm$9 & \\ % 6.847
\nodata & 1215.6701 & 3.0144$\pm$0.0001 & \phm{~}1140$\pm$10\phm{~~} & $-$22.03$\pm$1.17 & 151$\pm$8 & \\ % 6.847
\nodata & 1215.6701 & 3.0164$\pm$0.0004 & \phm{~~}470$\pm$10\phm{~~} & $-$4.72$\pm$0.86 & \phm{~}32$\pm$6 & \\ % 6.847
O\,V] & 1218.344 & 3.0204$\pm$0.0003 & \phm{~~}290$\pm$70\phm{~~}  & $-$0.37$\pm$0.11 & \phm{~~}3$\pm$1 & \\ % 6.825
N\,V   & 1240.14 & 3.0104$\pm$0.0002 & \phm{~}2230$\pm$20\phm{~~}  & $-$9.99$\pm$1.02 & \phm{~}66$\pm$7 & \\ % 6.656
\nodata   & 1240.14 & 3.0208$\pm$0.0002 & \phm{~~}700$\pm$10\phm{~~}  & $-$1.73$\pm$0.23 & \phm{~}12$\pm$2 & \\ % 6.656
\nodata   & 1240.14 & 3.0243$\pm$0.0004 & \phm{~}5400$\pm$70\phm{~~}  & $-$16.81$\pm$2.20 & 112$\pm$15 & \\ % 6.656
Si\,II & 1262.59 & 3.0239$\pm$0.0014 & \phm{~}2800$\pm$40\phm{~~}  & $-$3.36$\pm$0.47 & \phm{~}22$\pm$3 & \\ % 6.487
O\,I/Si\,II & 1305.42 & 3.0159$\pm$0.0042 & \phm{~}3000$\pm$200\phm{~} & $-$5.09$\pm$0.92 & \phm{~}32$\pm$6 & blend\\ % 6.185
C\,II  & 1335.30 & 3.0040$\pm$0.0016 & \phm{~}2200$\pm$300\phm{~} & $-$3.45$\pm$0.97 & \phm{~}21$\pm$6 & \\ % 5.989
Si\,IV/O\,IV] & 1398.33 & 2.9914$\pm$0.0061 & 10000$\pm$1300 & $-$6.74$\pm$0.90 & \phm{~}38$\pm$5 & blend\\ % 5.608
Si\,IV & 1393.755 & 3.0153$\pm$0.0008 & \phm{~}1000$\pm$200\phm{~} & $-$2.30$\pm$0.45 & \phm{~}13$\pm$3 & \\ % 5.634
\nodata & 1402.770 & 3.0143$\pm$0.0007 & \phm{~}1300$\pm$150\phm{~} & $-$3.30$\pm$0.39 & \phm{~}18$\pm$2 & \\ % 5.582
C\,IV  & 1549.06 & 2.9997$\pm$0.0010 & \phm{~}9000$\pm$100\phm{~} & $-$41.62$\pm$0.77 & 202$\pm$4 & \\ % 4.846
\nodata  & 1549.06 & 3.0031$\pm$0.0003 & \phm{~}2670$\pm$70\phm{~~}  & $-$13.04$\pm$0.36 & \phm{~}63$\pm$2 & \\ % 4.846
\nodata  & 1549.06 & 3.0178$\pm$0.0001 & \phm{~}1100$\pm$20\phm{~~}  & $-$14.59$\pm$0.34 & \phm{~}71$\pm$2 & \\ % 4.846
Fe\,II & \nodata & \nodata           & 13000$\pm$400\phm{~} & $-$12.38$\pm$0.39 & \phm{~}57$\pm$2 & \\ % 4.626
He\,II & 1640.42 & 3.0091$\pm$0.0006 & \phm{~}1200$\pm$80\phm{~~}  &  $-$2.06$\pm$0.12 & \phm{~~}9$\pm$1 & \\ % 4.466
O\,III] & 1663.48 & 3.0160$\pm$0.0015 &\phm{~}3200$\pm$150\phm{~} &  $-$3.18$\pm$0.16 & \phm{~}14$\pm$1 & \\ % 4.378
Al\,II  & 1721.89 & 3.0292$\pm$0.0018 & \phm{~}1500$\pm$400\phm{~} &  $-$0.87$\pm$0.22  & \phm{~~}4$\pm$1 & \\ % 4.168
N\,III] & 1750.26 & 3.0157$\pm$0.0014 & \phm{~}2000$\pm$300\phm{~} &  $-$1.63$\pm$0.27 & \phm{~~}7$\pm$1 & \\ % 4.072
Fe\,II  & 1788.73 & 3.0117$\pm$0.0012 & \phm{~}1900$\pm$250\phm{~} &  $-$2.18$\pm$0.28 & \phm{~~}9$\pm$1 & \\ % 3.947
Al\,III & 1857.40 & 3.0411$\pm$0.0074 & \phm{~}6300$\pm$1000 & $-$6.00$\pm$0.99 & \phm{~}22$\pm$4 & \\ % 3.741
Si\,III] & 1892.03 & 3.0133$\pm$0.0049 & \phm{~~}700$\pm$350\phm{~} &  $-$0.52$\pm$0.25 & \phm{~~}2$\pm$1 & \\ % 3.644
C\,III] & 1908.73 & 3.0131$\pm$0.0026 & \phm{~}5500$\pm$250\phm{~} & $-$19.74$\pm$1.33 & \phm{~}71$\pm$5 & \\ % 3.598
\nodata & 1908.73 & 3.0141$\pm$0.0004 & \phm{~~}900$\pm$70\phm{~~}  &  $-$4.39$\pm$0.32 & \phm{~}16$\pm$1 & \\ % 3.598
\enddata
\tablecomments{Emission-line fluxes and REW values have been corrected for
overlying associated absorption.
All errors are $\pm 1 \sigma$ statistical errors.}
\tablenotetext{a}{Vacuum rest wavelength in 
\AA\ \markcite{myj88}({Morton}, {York}, \& {Jenkins} 1988).  
We use the $\lambda_{\rm lab}$ wavelengths given in Table 2 of
\markcite{sdss73}{Vanden Berk} {et~al.} (2001), except for cases of
well-separated doublets and the two close blends noted in the table, 
for which we use the $\lambda_{\rm obs}$ wavelengths in the same reference.}
\tablenotetext{b}{Rest-frame FWHM in km s$^{-1}$.
The rest frame used is that of the quasar ($z=3.020$).
The FWHM values have not been corrected for the instrumental FHWM
of $\sim$150\,\kms.  For most emission lines this correction is negligible.}
\tablenotetext{c}{Rest-frame Equivalent Width in \AA.
The uncertainties are propagated from the fractional uncertainty on the width
or the peak of the Gaussian fit, whichever uncertainty is larger.
The rest frames are the same as those used for the FWHM calculation.}
\tablenotetext{d}{Flux of each Gaussian emission-line component,
in units of 10$^{-17}$ ergs cm$^{-2}$ s$^{-1}$.}
\tablenotetext{e}{Emission line blends are denoted `blend'.}
\end{deluxetable}

\begin{deluxetable}{llccrl}
\tablecaption{SDSS J0952+0114 Absorption Line Identifications\label{t_abs}}
%\tabletypesize{\footnotesize}
\tabletypesize{\scriptsize}
\tablewidth{450.00000pt}
\tablehead{
\colhead{Line}& \colhead{$\lambda_{rest}$\tablenotemark{a}}& \colhead{Redshift}&
\colhead{FWHM\tablenotemark{b}}& \colhead{REW\tablenotemark{c}}
&\colhead{Notes\tablenotemark{d}}
}
\startdata
\multicolumn{6}{l}{Associated Absorber A, $z=3.0097\pm0.0014$, $\Delta v=770\pm100$\,\kms}\\
\tableline
Ly$\gamma$ & \phm{1}972.5368 &	3.0097$\pm$0.0005 &  490$\pm$140 & 1.93$\pm$0.65 & \\
Ly$\beta$ & 1025.7223 &	3.0056$\pm$0.0006 &  980$\pm$100 & 2.92$\pm$0.30 &blend O\,I,Ly$\beta$(B)\\
O\,VI	& 1031.9265 &	3.0103$\pm$0.0002 &  240$\pm$30\phm{~} & 0.68$\pm$0.10 & \\
C\,II	& 1036.3367 &	3.0109$\pm$0.0002 &  650$\pm$30\phm{~} & 5.39$\pm$0.29 &blend\\
C\,II*	& 1037.0181 &	3.0083$\pm$0.0002 &  650$\pm$30\phm{~} & 5.39$\pm$0.29 &blend\\
O\,VI	& 1037.6155 &	3.0060$\pm$0.0002 &  650$\pm$30\phm{~} & 5.39$\pm$0.29 &blend\\
O\,I	& 1039.2304 &	3.0106$\pm$0.0003 &  140$\pm$20\phm{~} & 0.30$\pm$0.10 & \\
Si\,III	& 1206.500 &	3.0091$\pm$0.0014 &  290$\pm$20\phm{~} & 0.85$\pm$0.23 & \\
Ly$\alpha$ & 1215.6701 & 3.0084$\pm$0.0001 & 540$\pm$30\phm{~} & 8.31$\pm$0.53 & log($N_{H I}$)$\sim$18.7\\
N\,V	& 1238.821 &	3.0100$\pm$0.0008 &  230$\pm$10\phm{~} & 0.61$\pm$0.17 & \\
N\,V	& 1242.804 &	3.0091$\pm$0.0006 &  230$\pm$10\phm{~} & 0.24$\pm$0.09 &tentative\\ 
S\,II	& 1250.583 &	3.0075$\pm$0.0004 &  340$\pm$10\phm{~} & 0.38$\pm$0.07 & \\
S\,II	& 1253.808 &	3.0099$\pm$0.0005 &  188$\pm$2\phm{~~} & 0.23$\pm$0.03 & \\
S\,II	& 1259.519 &	3.0095$\pm$0.0002 &  144$\pm$2\phm{~~} & 0.23$\pm$0.04 & \\
Si\,II	& 1260.4223 &	3.0095$\pm$0.0004 &  216$\pm$2\phm{~~} & 0.54$\pm$0.11 & blend Fe\,II $\lambda$2374 ($z$=1.1291)\\
Si\,II*	& 1264.7377 &	3.0105$\pm$0.0007 &  220$\pm$10\phm{~} & 0.73$\pm$0.18 & blend Fe\,II $\lambda$2382 ($z$=1.1291)\\
O\,I	& 1302.1685 &	3.0094$\pm$0.0003 &  280$\pm$10\phm{~} & 1.28$\pm$0.47 & \\
Si\,II	& 1304.3711 &	3.0119$\pm$0.0004 &  330$\pm$10\phm{~} & 1.54$\pm$0.37 &blend\\
O\,I*	& 1304.8576 &	3.0104$\pm$0.0004 &  330$\pm$10\phm{~} & 1.54$\pm$0.37 &blend\\
O\,I*	& 1306.0286 &	3.0068$\pm$0.0004 &  330$\pm$10\phm{~} & 1.54$\pm$0.37 &blend\\
Si\,II*	& 1309.2757 &	3.0125$\pm$0.0006 &  283$\pm$4\phm{~~} & 0.46$\pm$0.08 & \\
C\,II	& 1334.5323 &	3.0117$\pm$0.0005 &  490$\pm$70\phm{~} & 1.71$\pm$0.25 &blend\\
C\,II*	& 1335.71 &	3.0082$\pm$0.0005 &  490$\pm$70\phm{~} & 1.71$\pm$0.25 &blend\\
Si\,IV	& 1393.755 &	3.0099$\pm$0.0001 &  180$\pm$30\phm{~} & 0.67$\pm$0.11 & \\
Si\,IV	& 1402.770 &	3.0102$\pm$0.0002 &  180$\pm$20\phm{~} & 0.55$\pm$0.08 & \\
Si\,II	& 1526.7071 &	3.0103$\pm$0.0002 &  180$\pm$10\phm{~} & 0.53$\pm$0.04 & \\
Si\,II*	& 1533.4312 &	3.0101$\pm$0.0002 &  170$\pm$20\phm{~} & 0.23$\pm$0.02 &blend C\,IV(D)\\
C\,IV	& 1548.202 &	3.0101$\pm$0.0001 &  230$\pm$10\phm{~} & 2.24$\pm$0.12 &blend C\,IV(B)\\
C\,IV	& 1550.774 &	3.0101$\pm$0.0001 &  230$\pm$10\phm{~} & 2.66$\pm$0.16 & \\
Fe\,II	& 1608.4511 &	3.0106$\pm$0.0002 &  200$\pm$40\phm{~} & 0.34$\pm$0.07 & \\
Al\,II	& 1670.7874 &	3.0103$\pm$0.0002 &  160$\pm$10\phm{~} & 0.41$\pm$0.04 & \\
Si\,II	& 1808.0126 &	3.0103$\pm$0.0003 &  120$\pm$50\phm{~} & 0.19$\pm$0.07 &tentative\\
Al\,III	& 1854.7164 &	3.0106$\pm$0.0002 &  \phm{~}70$\pm$20\phm{~} & 0.11$\pm$0.08 &tentative\\
Al\,III	& 1862.7865 &	3.0114$\pm$0.0005 &  170$\pm$60\phm{~} & 0.17$\pm$0.07 &tentative\\
Mg\,I	& 2026.4768 &	3.0101$\pm$0.0002 &  140$\pm$60\phm{~} & 0.28$\pm$0.13 &tentative\\
\tableline
\multicolumn{6}{l}{Associated Absorber B, $z=3.0025\pm0.0001$, $\Delta v=1310\pm10$\,\kms}\\
\tableline
Ly$\gamma$ & \phm{1}972.5368 &	3.0025$\pm$0.0005 &  270$\pm$40\phm{~}  & 0.94$\pm$0.32 & \\
Ly$\beta$ & 1025.7223 &	3.0056$\pm$0.0006 &  980$\pm$100  & 2.92$\pm$0.30 &blend Ly$\beta$(A)\\
O\,VI	& 1031.9265 &	3.0027$\pm$0.0003 &  100$\pm$20\phm{~}  & 0.10$\pm$0.03 & \\ 
C\,II	& 1036.3367 &	3.0010$\pm$0.0001 &  410$\pm$10\phm{~}  & 3.27$\pm$0.15 & \\
O\,VI	& 1037.6155 &	3.0060$\pm$0.0002 &  650$\pm$30\phm{~}  & 5.39$\pm$0.29 &blend C\,II*,C\,II(A),O\,VI(A)\\
Si\,III	& 1206.500 &	3.0028$\pm$0.0010 &  300$\pm$30\phm{~}  & 0.75$\pm$0.17 &blend Ly$\alpha$(D)\\
Ly$\alpha$ & 1215.6701 & 3.0025$\pm$0.0001 & 200$\pm$20\phm{~}  & 1.14$\pm$0.22 & log($N_{H I}$)$\sim$18\\
C\,II	& 1334.5323 &	3.0038$\pm$0.0011 &  620$\pm$100  & 1.40$\pm$0.26 &blend\\
C\,II*	& 1335.71 &	3.0003$\pm$0.0011 &  620$\pm$100  & 1.40$\pm$0.26 &blend\\
C\,IV	& 1548.202 &	3.0020$\pm$0.0001 &  110$\pm$10\phm{~}  & 0.12$\pm$0.01 & \\
C\,IV	& 1550.774 &	3.0034$\pm$0.0001 &  230$\pm$10\phm{~}  & 2.24$\pm$0.12 &blend C\,IV(A)\\
\tableline
\multicolumn{6}{l}{Associated Absorber C, $z=2.9926\pm0.0025$, $\Delta v=2050\pm190$\,\kms}\\
\tableline
Ly$\alpha$ & 1215.6701 & 2.9926$\pm$0.0025 & 200$\pm$20\phm{~}  & 0.44$\pm$0.12 & log($N_{H I}$)$\sim$18\\
\tableline
\multicolumn{6}{l}{Associated Absorber D, $z=2.9724\pm0.0004$, $\Delta v=3570\pm20$\,\kms}\\
\tableline
O\,VI	& 1031.9265 &	2.9701$\pm$0.0003 &  220$\pm$70\phm{~}  & 0.14$\pm$0.06 &tentative\\
O\,VI	& 1037.6155 &	2.9725$\pm$0.0006 &  480$\pm$120 & 0.72$\pm$0.18 & blend Ly$\beta$(A)\\
Ly$\alpha$ & 1215.6701 & 2.9726$\pm$0.0010 & 300$\pm$30\phm{~}  & 0.75$\pm$0.17 &blend Si\,III(B); log($N_{H I}$)$\lesssim$18.2\\
C\,IV	& 1548.202 &	2.9719$\pm$0.0002 &  160$\pm$20\phm{~}  & 0.23$\pm$0.02 &blend Si\,II*(A)\\
C\,IV	& 1550.774 &	2.9728$\pm$0.0007 &  170$\pm$20\phm{~}  & 0.14$\pm$0.01 & \\
\tableline
\multicolumn{6}{l}{Associated Absorber E, $z=2.9661\pm0.0005$, $\Delta v=4050\pm40$\,\kms}\\
\tableline
Ly$\beta$ & 1025.7223 &	2.9663$\pm$0.0005 &  500$\pm$110 & 0.63$\pm$0.14 &blend, Ly$\alpha$ forest\\
Ly$\alpha$ & 1215.6701 & 2.9662$\pm$0.0004 & 190$\pm$20\phm{~} & 0.12$\pm$0.03 &log($N_{H I}$)$\sim$18\\
N\,V	& 1238.821 &	2.9667$\pm$0.0010 &  250$\pm$20\phm{~} & 0.25$\pm$0.02 & \\
N\,V	& 1242.804 &	2.9659$\pm$0.0006 &  210$\pm$50\phm{~} & 0.19$\pm$0.04 & \\
\tableline
\multicolumn{6}{l}{Intervening Absorber, $z=2.7713\pm0.0002$}\\
\tableline
C\,IV	& 1548.202 &	2.7711$\pm$0.0002 &  200$\pm$20\phm{~} & 0.28$\pm$0.03 & \\
C\,IV	& 1550.774 &	2.7721$\pm$0.0004 &  270$\pm$10\phm{~} & 0.21$\pm$0.02 & \\
\tableline
\multicolumn{6}{l}{Intervening Absorber, $z=1.1291\pm0.0005$}\\
\tableline
Fe\,II	& 2249.88   &	1.1291$\pm$0.0008 &  210$\pm$60\phm{~} & 0.27$\pm$0.12 &blend? Ly$\alpha$ forest\\
Fe\,II	& 2260.7805 &	1.1287$\pm$0.0005 &  300$\pm$40\phm{~} & 0.31$\pm$0.05 &blend? Ly$\alpha$ forest\\
Fe\,II	& 2344.2139 &	1.1287$\pm$0.0002 &  160$\pm$30\phm{~} & 0.16$\pm$0.03 & \\
Fe\,II	& 2374.4612 &	1.1289$\pm$0.0002 &  \phm{~}90$\pm$20\phm{~} & 0.19$\pm$0.04 & blend Si\,II(A)\\
Fe\,II	& 2382.7652 &	1.1287$\pm$0.0004 &  230$\pm$60\phm{~} & 0.73$\pm$0.18 & blend Si\,II*(A)\\
Fe\,II	& 2586.65 &	1.1301$\pm$0.0002 & 310$\pm$140 & 0.35$\pm$0.16 & \\
Fe\,II	& 2600.1729 &	1.1293$\pm$0.0002 &  210$\pm$50\phm{~} & 0.39$\pm$0.09 & \\
Mg\,II	& 2796.352 &	1.1287$\pm$0.0001 &  310$\pm$30\phm{~} & 0.66$\pm$0.05 & \\
Mg\,II	& 2803.531 &	1.1286$\pm$0.0001 &  300$\pm$20\phm{~} & 0.45$\pm$0.03 & \\
Mg\,I	& 2852.9642 &	1.1296$\pm$0.0001 &  330$\pm$10\phm{~} & 0.37$\pm$0.02 & \\
\enddata
\tablecomments{All errors are $\pm 1 \sigma$ statistical errors.}
\tablenotetext{a}{Vacuum rest wavelength in 
\AA\ \markcite{myj88}({Morton}, {York}, \& {Jenkins} 1988).}
\tablenotetext{b}{Rest-frame FWHM in km s$^{-1}$.  For associated
absorption lines, the rest frame used is that of the quasar ($z=3.020$).
For intervening absorption line systems, the rest frame redshift is listed in
the table.  The FWHM values have not been corrected for the instrumental FHWM
of $\sim$150\,\kms.  Absorption lines with measured 
FWHM$\lesssim$150\,\kms\ indicate detections of borderline significance.}
\tablenotetext{c}{Rest-frame Equivalent Width in \AA.  
The uncertainties are propagated from the fractional uncertainty on the width 
or the peak of the Gaussian fit, whichever uncertainty is larger.
The rest frames are the same as those used for the FWHM calculation.}
\tablenotetext{d}{Successive lines in the table denoted `blend' are blended with
each other.  Blends between lines in different absorbers are also indicated.
No attempt at separation has been made for either type of blend; naturally,
the FWHM values for these blends are usually larger than for unblended lines.
For Ly$\alpha$ lines, {\em approximate} log($N_{H I}$) values are given in
cm$^{-2}$; see \S\,\ref{FIT}.
}
\end{deluxetable}

\end{document}